\documentclass{Interspeech2024}
\usepackage{ctable}
\usepackage{multirow}
\usepackage{amsmath}
\usepackage{amssymb}
\usepackage{cite}
\usepackage[utf8]{inputenc}
\usepackage{setspace}
\usepackage{lipsum}
\usepackage{caption}
\usepackage[subtle]{savetrees}

\usepackage{subcaption}
\usepackage{booktabs}
\usepackage{hyperref}
\usepackage{enumerate}
\usepackage{hyperref}
\usepackage[utf8]{inputenc}

\usepackage{mathtools}
\DeclarePairedDelimiter{\norm}{\lVert}{\rVert}

\newcommand{\newpara}[1]{\vspace{3pt}\noindent\textbf{#1}}

\DeclarePairedDelimiter{\abs}{\lvert}{\rvert}

\interspeechcameraready

\title{Disentangled Representation Learning \\for Environment-agnostic Speaker Recognition}

\name[affiliation={1}]{KiHyun}{Nam}
\name[affiliation={2}]{Hee-Soo}{Heo}
\name[affiliation={3}]{Jee-weon}{Jung}
\name[affiliation={1}]{Joonson}{Chung}

\address{
  $^1$Korea Advanced Institute of Science and Technology, South Korea\\
  $^2$Naver Cloud Corporation, South Korea \\
  $^3$Carnegie Mellon University, USA}
\email{nkh.mmai@kaist.ac.kr, heesoo.heo@navercorp.com, jeeweonj@ieee.org, joonsc@kaist.ac.kr}


\begin{document}

\maketitle

\begin{abstract}
This work presents a framework based on feature disentanglement to learn speaker embeddings that are robust to environmental variations.
Our framework utilises an auto-encoder as a disentangler, dividing the input speaker embedding into components related to the speaker and other residual information. 
We employ a group of objective functions to ensure that the auto-encoder's code representation -- used as the refined embedding -- condenses only the speaker characteristics.
We show the versatility of our framework through its compatibility with any existing speaker embedding extractor, requiring no structural modifications or adaptations for integration. 
We validate the effectiveness of our framework by incorporating it into two popularly used embedding extractors and conducting experiments across various benchmarks.
The results show a performance improvement of up to 16\%.
We release our code for this work to be available here\footnote{Official webpage: \href{https://mm.kaist.ac.kr/projects/voxceleb-disentangler/}{{https://mm.kaist.ac.kr/projects/voxceleb-disentangler/}}

\hspace{4.5pt} Official code: https://github.com/kaistmm/voxceleb-disentangler}.

\end{abstract}

\section{Introduction}

The growth of voice-based AI services has amplified the demand for robust speaker recognition models capable of operating effectively in noisy environments. Every audio recording carries not only the speaker-specific  characteristics~\cite{luu21_interspeech,luu2022investigating}, such as age, gender, accent~\cite{raj2019probing},  emotion~\cite{williams2019disentangling, pappagari2020x} and language~\cite{maiti2020generating, nam2022disentangled}, but also environmental information~\cite{campbell1997speaker} like noise and reverberation. These factors are intertwined within the speaker representation. While some of these factors are essential for identifying the speaker, others, particularly environmental information, can act as obtrusive information, making speaker recognition more challenging. This issue becomes more pronounced in an environment mismatch scenario, where changes in recording conditions -- ranging from serene offices to bustling streets -- can drastically alter audio characteristics, to the extent that they may seem to originate from different individuals. Consequently, the removal of these intrusive factors from speaker embeddings emerges as a pivotal step towards enhancing the effectiveness of speaker recognition systems, ensuring that they can distinguish between essential speaker-related information and environmental distortions.

Despite the use of datasets~\cite{nagrani2020voxceleb, chung18b_interspeech} recorded in various real-world environments and data augmentation techniques~\cite{snyder2015musan, ko2017study}, the emergence of realistic benchmark datasets~\cite{nagrani2020voxsrc,brown2021playing, huh2023voxsrc} continually reveals the vulnerability of speaker recognition systems in varied environment conditions. This highlights the need for a more fundamental solution that can explicitly exclude environmental information from speaker representations.

Disentangled representation learning (DRL)~\cite{wang2022disentangled} emerges as a reasonable approach for tackling this challenge. DRL seeks to independently isolate and manipulate the different factors within the input data, showing promise across various fields~\cite{luu2022investigating, sang2021deaan, JangOC0CK22, nam2022disentangled, yang2022disentangled, omran2022disentangling, hao2023learning}. For example, \cite{nam2022disentangled} removes linguistic information from speaker representation using an adversarial-based DRL framework for bilingual speaking scenarios. DRL represents a promising approach, yet its occasional removal of vital information, which can compromise performance, highlights the need for continued refinement and exploration.

We propose a novel adversarial learning-based disentangled representation learning framework that can remove environmental information from speaker representations while minimising the loss of speaker information. 
Traditional adversarial learning-based DRL ensures effective information removal, but adversarial learning often distorts task-relevant information, resulting in training instability~\cite{arjovsky2017towards, kang2020disentangled, wang2022disentangledantisoof, nam2022disentangled}. 
We introduce a new idea that uses an auto-encoder as a disentangler to separate environmental information while leveraging a reconstruction loss function of the auto-encoder to penalise unnecessary information loss, thereby mitigating the loss of vital speaker information during the DRL process. Additionally, we combine a set of objective functions to facilitate the learning of refined speaker information within the disentangled speaker embedding. To assist environment-DRL, we employ a regularisation technique that swaps embeddings of the same speaker from different environmental origins during the reconstruction process. Another contribution of our framework is the adaptability to seamlessly integrate with various existing speaker recognition networks without any structural modifications. Experimental results show that our framework demonstrates up to 16\% performance improvement over baseline models and previous DRL framework on various evaluation sets.

\begin{figure*}[t]
  \centering
  \includegraphics[width=1\linewidth]{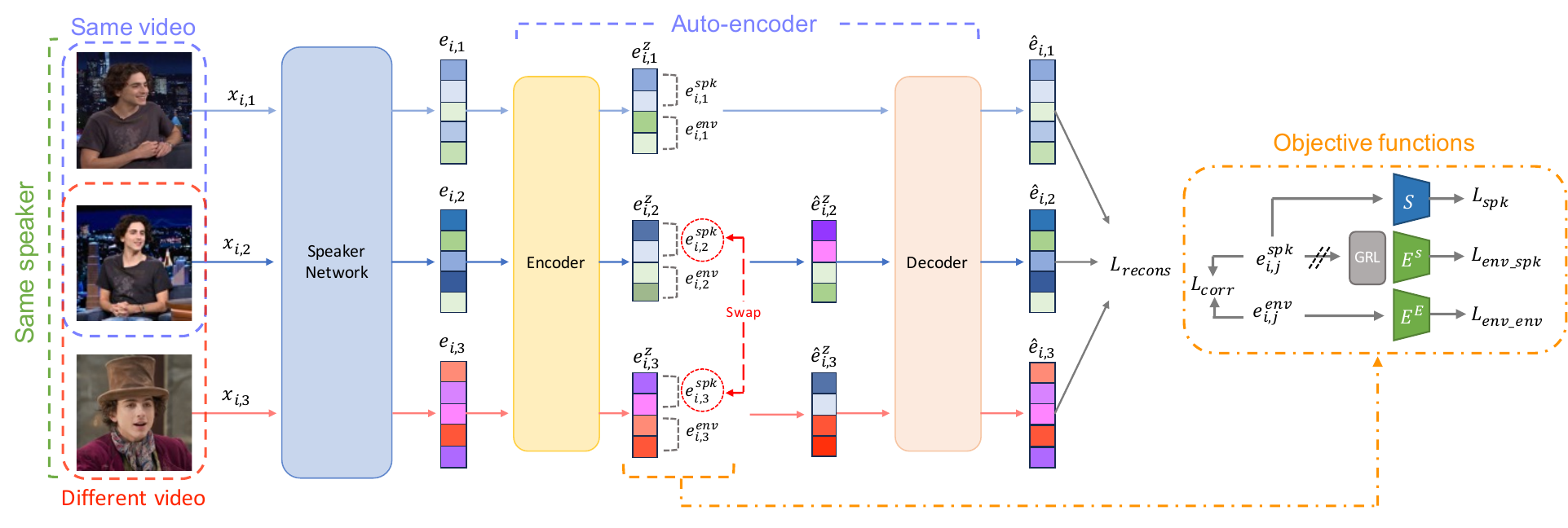}
    \vspace{-0.7cm}
  \caption{The illustration of the proposed environment-disentangled representation learning framework. Auto-encoder encodes the speaker network's entangled speaker representation into a compact latent vector, which is then divided into distinct speaker and environment representation vectors. Orange box represents a set of objective functions to facilitate the learning of refined speaker and environment representations from the auto-encoder's bottleneck representation. Reconstruction training of the auto-encoder minimises the loss of vital speaker information during the disentangled representation learning.}
  \label{fig:framework_figure}
  \vspace{-0.3cm}
\end{figure*}

We summarise our contributions as follows: (1) We introduce a novel DRL framework, which leverages an auto-encoder as a disentangler, to minimise the loss of vital information. (2) Our framework is easily adaptable to existing speaker networks without any structural modifications. (3) Our framework shows significant performance improvements on various evaluation sets reflecting in the wild conditions and also increases the performance of existing baseline models on standard benchmarks.

\section{Related works}
\newpara{Triplet batch formulation.}
Our batch formulation is similar to those found in previous studies~\cite{chung2019delving, kang2022aat}, which employed a triplet batch formulation -- that is, each mini-batch index comprises three utterances. 
\cite{chung2019delving} constructed triplets of utterances from the same speaker by selecting two utterances from the same video and the third from a different, non-overlapping video. 
However, this approach did not incorporate further data augmentation, ensuring that the first two utterances were subjected to similar environmental noises, whereas the third was exposed to distinct noises. \cite{kang2022aat} also adopted a triplet batch formulation strategy but without utilising video session information, extracting all three non-overlapping utterances from a single video and then simulating an environment mismatch scenario through artificial data augmentation.

Our strategy takes advantage of the studies above. We assemble the triplets in the same manner as \cite{chung2019delving} and additionally implement data augmentation techniques akin to those used by \cite{kang2022aat}.

\newpara{Feature enhancement using auto-encoder.}
Auto-encoders have been widely used to enhance latent embeddings for various purposes~\cite{kim2023advancing, omran2022disentangling, JangOC0CK22}.
\cite{kim2023advancing} introduced a method for enhancing speaker embeddings using an auto-encoder to reduce noise in speaker diarisation. However, this method did not specifically address the disentanglement of environmental noise, and the auto-encoder training was performed online within a diarisation framework using a single input recording. 
\cite{omran2022disentangling,JangOC0CK22} introduce a disentangling auto-encoder to reduce background information within audio signal and sign language embeddings. These studies proposed an embedding swapping technique to enhance the disentanglement capability.

Our work introduces an auto-encoder-based DRL framework that, for the first time, targets the disentanglement of environmental noise in speaker verification.
Our proposed framework utilises an auto-encoder combined with a variety of objective functions.
The framework incorporates adversarial learning and embedding swapping techniques designed to accurately disentangle environmental noise while preserving the speaker's fundamental characteristics.

\section{Proposed disentanglement framework}
This section presents the proposed environment DRL framework, as illustrated in Figure 1.
Employing a set of triplets obtained through a specialised batch sampling method (Section 3.1), the auto-encoder condenses and reconstructs the input embeddings, as detailed in Section 3.2 ($L_{recons}$).
Simultaneously, four additional objective functions ($L_{env\_env}$, $L_{env\_spk}$, $L_{corr}$, and $L_{spk}$) facilitate the training process to effectively isolate environmental noise while retaining essential speaker features, as discussed in Sections 3.3 and 3.4.

\subsection{Batch construction with data augmentation}

Our batch formulation is essentially identical to that of \cite{chung2019delving}.
Each mini-batch index consists of three utterances: $ x_{i,1}$, $ x_{i,2}$, and $ x_{i,3}$, where $i$ denotes the mini-batch index. All three utterances originate from the same speaker; however, the first two are sourced from an identical video, and the third is drawn from a different video. 
This setup aims to ensure that the first two utterances reflect the same environmental conditions, whereas the third introduces a distinct environment. 

The novelty in batch construction stems from data augmentation. In contrast to \cite{chung2019delving}, we apply identical augmentation techniques to the first two utterances and a different augmentation method to the third. This further ensures that the first two utterances share similar environmental noise while the third utterance involves distinct environmental noise.

\subsection{Framework structure and reconstruction}
Our disentanglement framework is independent of the speaker embedding extractor and corresponds to the block denoted as ``Auto-encoder'' in Figure 1. The framework inputs an arbitrary extractor's embedding, \(e_{i,j} \in \mathbb{R}^{D}\)where $e_{i,j}$ denotes the extracted speaker embedding from an input utterance $x_{i,j}$, $j \in \{1, 2, 3\}$. 
The encoder of the auto-encoder projects the input speaker representation \(e_{i,j}\) into a compact representation \(e^z_{i,j}\). The decoder reconstructs \(\hat{e}_{i,j}\) using \(e^z_{i,j}\) and a reconstruction loss calculates the L1 distance between the output $\hat{e}_{i,j}$ and the input \(e_{i,j}\).
The reconstruction loss \(L_{recons}\) is formulated as follows:
 \begin{equation}
     L_{recons} = \sum_{j=1}^{3}(\abs{e_{i,j} - \hat{e}_{i,j}}).
\end{equation}
On top of the basic reconstruction loss of an auto-encoder introduced above, our framework employs several additional techniques and objective functions including adversarial learning loss to successfully train the model without collapsing or deteriorating.

\subsection{Code swapping between different environments}
Within the auto-encoder, the code \(e^z_{i,j}\) is further split into a speaker representation \(e_{i,j}^{spk} \in \mathbb{R}^{d}\) and an environment representation \(e_{i,j}^{env} \in \mathbb{R}^{D-d}\) as shown in the middle part of Figure 1. 
Among a pair of triplets, we swap $e^{spk}_{i,2}$ with $e^{spk}_{i,3}$. This process guides the model to condense core speaker characteristics in $e^{spk}_{i,j}$ while environmental noise is projected to $e^{env}_{i,j}$. 
Note that we do not apply code swapping to $e^z_{i,1}$

\subsection{Discriminator training}

To learn fine-grained speaker and environment information within the two outputs from the encoder, \(e^{spk}_{i,j}\) and \(e^{env}_{i,j}\), respectively, we train a total of three discriminators: A speaker discriminator \(S\) and two environment discriminators \(E^{E}\) and \(E^{S}\). 

The speaker discriminator \(S\) computes the speaker classification loss \(L_{spk}\) from \(e^{spk}_{i,j}\) to perform speaker recognition training. For \(L_{spk}\), we employ a combination loss function, which is proposed in \cite{kwon2021ins}, with an angular prototypical loss~\cite{chung20b_interspeech} and a vanilla softmax loss. For the \(M\) value of the angular prototypical loss~\cite{chung20b_interspeech}, we use \(M=3\); one sample  \(e^{spk}_{i,1}\) from \(i\)-th triplet as a query set and other two samples \(e^{spk}_{i,2}\) and \(e^{spk}_{i,3}\) as a support set. For the vanilla softmax loss, the speaker discriminator \(S\) uses one fully-connected layer \(f\) to map \(e^{spk}_{i,j}\) to a speaker class vector.

The environment discriminator \(E^{E}\) computes the environment loss function \(L_{env\_env}\) by passing the environment-specific representation to an environment classifier \(g\) for environment recognition training. For the environment loss function, we employ a triplet loss as follows:

\vspace{-0.2cm}
\begin{equation}\label{eqn:loss_env}
    \begin{gathered}
        pos\_dist = \norm*{g(e^{env}_{i,1}) - g(e^{env}_{i,2})}_2^2 \\
        neg\_dsit = \norm*{g(e^{env}_{i,1}) - g(e^{env}_{i,3})}_2^2 \\
        L_{env\_env} = max(0, m + pos\_dist - neg\_dist)
    \end{gathered}
\end{equation}

\noindent where \(m\) is the margin of the triplet loss. The environment loss function ensures that the environment discriminator develops the ability to distinguish environment representations, making similar environment representations from the same video closer and those from different videos further apart.

\subsection{Adversarial learning}
\vspace{-0.2cm}
The environment discriminator \(E^S\) performs to capture any residual environmental information from the disentangled speaker representation \(e^{spk}_{i,j}\). \(E^S\) uses the same network structure and loss function as \(E^E\), but they not share any parameters. \(L_{\text{env\_spk}}\) is equal to Eq. \ref{eqn:loss_env}, except that \(L_{\text{env\_spk}}\) replaces the input \(e^{env}_{i,j}\) with the \(e^{spk}_{i,j}\). Since \(E^S\) should be trained independently, \(E^S\) is not connected to other neural networks, and the gradient is not propagated below the input data \(e^{spk}_{i,j}\).

To explicitly remove residual environmental information from the speaker representation \(e^{spk}_{i,j}\), we employ a combination of the gradient reversal layer (GRL) with the correlation minimisation loss proposed by \cite{nam2022disentangled}. The GRL is attached in front of the environment discriminator \(E^S\), which inverts the gradient of the loss function \(L_{env\_spk}\), interfering with loss minimisation. By the GRL, our framework learns the speaker representation \(e^{spk}_{i,j}\) that disrupts the environment discriminator \(E^S\), thereby inducing the removal of the residual environmental information. We denote the loss function \(L_{env\_spk}\) passing through the GRL as \(L_{\text{env\_spk(G)}}\). As an additional regularisation, we employ the mean absolute pearson correlation (MAPC) loss, used in~\cite{kang2020disentangled, nam2022disentangled}, to minimise the correlation between speaker and environment representations. This loss function is denoted as \(L_{corr}\) and is formulated as follows:

\vspace{-0.15cm}
\newcommand{\Cov}{\operatorname{Cov}}
\begin{equation}
    L_{corr} = \frac{|\Cov({e^{spk}_{i,j}},{e^{env}_{i,j}})|}{\sigma({e^{spk}_{i,j}})\cdot \sigma({e^{env}_{i,j}})}
\end{equation}
\vspace{-0.5cm}
\mbox{}\\

\noindent where \(\Cov(\cdot)\) and \(\sigma(\cdot)\) mean the covariance and the standard deviation, respectively.

\noindent In summary, the overall loss function is defined as follows:

\vspace{-0.2cm}
\begin{equation}
    \begin{gathered}
        L_{total} = \lambda_{S} * L_{spk} + \lambda_{R} * L_{recons} + \lambda_{E} * L_{env\_env} \\ 
                  + \lambda_{adv} * L_{env\_spk(G)} + \lambda_{C} * L_{corr}
    \end{gathered}
\end{equation}

\noindent where the lambda values are the weight values of losses summation. Same as the training process of~\cite{nam2022disentangled}, for the same mini-batch, $L_{total}$ updates the framework modules excluding the environment discriminator \(E^S\) and $L_{env\_spk}$ updates only the \(E^S\).

\begin{table*}
\centering
\caption{EER and minDCF on (a) VoxSRC22\&23 and VC-Mix evaluation sets, (b) VoxCeleb1-based evaluation sets. All experiments are repeated three times, and we report the mean and the standard deviation. \textbf{GRL}~\cite{nam2022disentangled}: a prior work of the adversarial learning-based DRL framework using the gradient reversal layer;}
\vspace{-3mm}
\begin{subtable}[t]{0.8\textwidth}
\resizebox{1\linewidth}{!}{
\begin{tabular}[t]{@{\extracolsep{8pt}}lcccccccc@{}}
\toprule
\multirow{2}{*}{\bf Model} & \multicolumn{2}{c}{\bf VoxSRC22} & \multicolumn{2}{c}{\bf VoxSRC23} & \multicolumn{2}{c}{\bf VC-Mix}  \\
\cmidrule[0.2pt]{2-3} \cmidrule[0.2pt]{4-5} \cmidrule[0.2pt]{6-7} \cmidrule[0.2pt]{8-9}
& {\bf EER (\%)} & {\bf minDCF} & {\bf EER (\%)} & {\bf minDCF} & {\bf EER (\%)} & {\bf minDCF} \\
\toprule
\toprule

ResNet-34~\cite{kwon2021ins} &
3.25~$\pm$~0.041 & 0.211~$\pm$~0.0013 &
5.91~$\pm$~0.096 & 0.323~$\pm$~0.0028 & 
3.05~$\pm$~0.091 & 0.245~$\pm$~0.0051 \\
\hspace{3mm} {\it + GRL}~\cite{nam2022disentangled}  & 
3.15~$\pm$~0.101 & 0.209~$\pm$~0.0086 &
5.60~$\pm$~0.130 & 0.314~$\pm$~0.0062 & 
2.95~$\pm$~0.207 & 0.253~$\pm$~0.0157 \\
\hspace{3mm} {\it + \textbf{Ours}}  &
\textbf{2.95~$\pm$~0.047} & \textbf{0.193~$\pm$~0.0067} &
\textbf{5.35~$\pm$~0.126} & \textbf{0.306~$\pm$~0.0024} & 
\textbf{2.58~$\pm$~0.113} & \textbf{0.223~$\pm$~0.0132} \\
\toprule
ECAPA-TDNN~\cite{desplanques2020ecapa}&  
3.25~$\pm$~0.038 & 0.210~$\pm$~0.0015 & 
5.92~$\pm$~0.016 & 0.328~$\pm$~0.0018 & 
2.92~$\pm$~0.145 & 0.254~$\pm$~0.0057 \\
\hspace{3mm} {\it + GRL}~\cite{nam2022disentangled} & 
3.22~$\pm$~0.130 & 0.203~$\pm$~0.0074 & 
5.85~$\pm$~0.105 & \textbf{0.297~$\pm$~0.0056} & 
2.62~$\pm$~0.131 & \textbf{0.211~$\pm$~0.0089} \\
\hspace{3mm} {\it + \textbf{Ours}} & 
\textbf{3.11~$\pm$~0.065} & \textbf{0.199~$\pm$~0.0075} &
\textbf{5.81~$\pm$~0.090} & 0.325~$\pm$~0.0006 & 
\textbf{2.43~$\pm$~0.059} & 0.212~$\pm$~0.0008 \\
\bottomrule
\end{tabular}}
\caption{Results on VoxSRC22\&23 evaluation sets and VC-Mix.}
\label{tab:table2_a}
\end{subtable}

\vspace{-1.5mm}

\begin{subtable}[t]{0.8\linewidth}
\resizebox{1\linewidth}{!}{
\begin{tabular}[t]{ @{\extracolsep{8pt}}lcccccccc@{}}

\toprule
\multirow{2}{*}{\bf Model} & \multicolumn{2}{c}{\bf Vox1-O} & \multicolumn{2}{c}{\bf Vox1-E} & \multicolumn{2}{c}{\bf Vox1-H}  \\
\cmidrule[0.2pt]{2-3} \cmidrule[0.2pt]{4-5} \cmidrule[0.2pt]{6-7} \cmidrule[0.2pt]{8-9}
& {\bf EER (\%)} & {\bf minDCF} & {\bf EER (\%)} & {\bf minDCF} & {\bf EER (\%)} & {\bf minDCF} \\
\toprule
\toprule
ResNet-34~\cite{kwon2021ins} &
0.95~$\pm$~0.051 & 0.076~$\pm$~0.0048 &
1.26~$\pm$~0.028 & 0.089~$\pm$~0.0022 & 
2.51~$\pm$~0.038 & 0.162~$\pm$~0.0007 \\
\hspace{3mm} {\it + GRL}~\cite{nam2022disentangled}  & 
1.13~$\pm$~0.053 & 0.083~$\pm$~0.0078 &
1.16~$\pm$~0.035 & 0.081~$\pm$~0.0019 & 
2.34~$\pm$~0.033 & 0.153~$\pm$~0.0021 \\
\hspace{3mm} {\it + \textbf{Ours}}  &
\textbf{0.86~$\pm$~0.024} & \textbf{0.068~$\pm$~0.0047} &
\textbf{1.10~$\pm$~0.010} & \textbf{0.078~$\pm$~0.0015} & 
\textbf{2.20~$\pm$~0.016} & \textbf{0.142~$\pm$~0.0031} \\
\toprule
ECAPA-TDNN~\cite{desplanques2020ecapa} &  
0.89~$\pm$~0.024 & 0.072~$\pm$~0.0093 & 
1.16~$\pm$~0.003 & 0.081~$\pm$~0.0006 & 
2.39~$\pm$~0.003 & \textbf{0.153~$\pm$~0.0006} \\
\hspace{3mm} {\it + GRL}~\cite{nam2022disentangled} & 
0.90~$\pm$~0.046 & 0.076~$\pm$~0.0030 & 
1.19~$\pm$~0.025 & 0.083~$\pm$~0.0022 & 
2.51~$\pm$~0.035 & 0.163~$\pm$~0.0024 \\
\hspace{3mm} {\it + \textbf{Ours}} & 
\textbf{0.82~$\pm$~0.006} & \textbf{0.067~$\pm$~0.0016} &
\textbf{1.16~$\pm$~0.011} & \textbf{0.080~$\pm$~0.0021} & 
\textbf{2.38~$\pm$~0.007} & 0.156~$\pm$~0.0006 \\

\bottomrule
\end{tabular}}
\caption{Results on VoxCeleb1-based evaluation sets.}
\label{tab:table2_b}
\end{subtable}

\vspace{-7.5mm}
\label{tab:results_main}
\end{table*}

\section{Experiments}

\subsection{Input representations}
First, we randomly extract a 2-second audio segment from each utterance and apply pre-emphasis with a coefficient of 0.97. We transform the input signal into a spectrogram with a 25ms window size, 10ms stride size, and a hamming window and use it as input data for the speaker network. For the ResNet-34, we use a 64-dimensional log mel-spectrogram as the input data and applied instance normalisation~\cite{ulyanov2016instance} to the input. For the ECAPA-TDNN, we use 80-dimensional log mel-spectrogram as the input.

\subsection{Model architecture}

To demonstrate the compatibility of the proposed method on existing speaker recognition systems, we employ two existing models, the variant of ResNet-34 proposed in \cite{kwon2021ins} and ECAPA-TDNN \cite{desplanques2020ecapa}. Note in this paper, we do not use the residual fully-connected layers following the pooling layer of both speaker networks. To compare our framework with the prior adversarial learning-based DRL model, we employ the GRL-based framework proposed in~\cite{nam2022disentangled}.

\newpara{ResNet-34.} We choose `\textbf{H / ASP}' version model mentioned in \cite{kwon2021ins}, which uses the attentive statistic pooling (ASP) \cite{okabe18_interspeech} layer.

\newpara{ECAPA-TDNN.} ECAPA-TDNN \cite{desplanques2020ecapa} is a neural network, which consists of a series of 1-dimensional Res2Blocks. ECAPA-TDNN uses the channel- and context-dependent statistics pooling layer. We employ the large-size model used in \cite{desplanques2020ecapa}.

\newpara{Auto-encoder.} The auto-encoder's encoder and decoder each consist of one batch normalisation layer followed by one fully-connected layer, sequentially. The input and output dimension sizes of the encoder and decoder are symmetrical. For the encoder, the input dimension size matches the output size of the speaker network's pooling layer, while the dimension size of the output vector \(e^z\) is 1024 and 512 for ResNet-34 and ECAPA-TDNN, respectively. The latent representation \(e^z\) is divided in half, equally split into the \(e^{spk}\) and the \(e^{env}\). Before being passed to the decoder, we use L1 normalisation to both \(e^{spk}\) and \(e^{env}\) independently.

\newpara{Discriminator.} For the speaker discriminator \(S\), we use just one fully-connected layer as \(f\) for the cross-entropy loss. Therefore, the output dimension size of the \(f\) is the same as the number of speaker classes. For the \(g\) of the environment discriminators \(E^E\) and \(E^S\), we use two MLP layers and each MLP layer consists of a batch normalisation layer, an ELU~\cite{clevert2015fast} activation function, and a fully-connected layer, sequentially. For ResNet-34 and ECAPA-TDNN, the output sizes of the first MLP layer are 512 and 256, respectively, and the output sizes of the last MLP layer are 512 and 128, respectively.

\subsection{Implementation details}

\newpara{Datasets.} For training, we use the development sets from VoxCeleb2~\cite{chung18b_interspeech}. Since the VoxCeleb datasets provide video session information for each speaker, we can utilise the video session information for the batch configuration described in Section 2.1. For evaluation, we select 6 multiple evaluation sets: three evaluation protocols utilising VoxCeleb1~\cite{nagrani2020voxceleb}, VoxSRC22~\cite{huh2023voxsrc}\ and 23, and VC-Mix~\cite{heo2023rethinking}. VoxSRC 22\&23 and VC-Mix are the evaluation sets that reflect the environment mismatch problem. For data augmentation, we employ reverberations of simulated RIRs dataset~\cite{ko2017study} and noises from the MUSAN dataset~\cite{snyder2015musan}.

\newpara{Training.}
All experiments are based on the PyTorch framework~\cite{paszke2019pytorch} and open-source \texttt{voxceleb\_trainer}\footnote{\url{https://github.com/clovaai/voxceleb_trainer}}. We use mixed precision training and the Adam Optimizer~\cite{kingma20153rd} with an initial learning rate of 0.001. ResNet-34-based experiments have a batch size of 220 and reduce the learning rate by 25\% every 16 epochs. ECAPA-TDNN-based experiments have a batch size of 256 and decrease the learning rate by 25\% every 8 epochs. Our implementation is performed on a single NVIDIA RTX 4090 GPU with 24 GB memory. Only the value of \(\lambda_{adv}\) is 0.5 and the values of other \(\lambda\) are 1. The training takes around 300 epochs. For all statistic pooling layers in the baseline and our models, all mean pooling parts are replaced with \(\l_{2}\)-norm pooling~\cite{wang2021revisiting}.

\newpara{Evaluation.}
We use three datasets, VoxSRC22\&23 and VC-Mix, as evaluation sets to measure the environment robustness performance. Since the three datasets, Vox1-O, Vox1-E, and Vox1-H, are not designed to be dependent on a specific factor, we evaluate the generalisation performance of these three datasets. We measure the performances by two metrics: 1) the Equal Error Rate (EER), where the rates of False Rejections (FRR) and False Alarms (FAR) are identical, and 2) the minimum Detection Cost Function (minDCF) described in NIST SRE~\cite{omid19nsre}, which is a weighted sum of FRR and FAR. For minDCF, we use the parameters \(C_{miss} = 1,\ C_{fa}=1\) and \(P_{target} = 0.05\). We sample each utterance into ten segments of 4 seconds each and compute the similarity across all segment pair combinations. The average similarity score is used as the trial's final score. This scoring is mentioned in~\cite{kwon2021ins}.

\section{Results}
Our experimental results are summarised in Table \ref{tab:results_main}. We compare three versions of each model, baseline, using only GRL~\cite{nam2022disentangled} and using our framework. For reliable measurements, we train all models three times with different random seeds and report the mean and the standard deviations. We use the standard deviation values to measure the training stability of models. Table \ref{tab:table2_a} demonstrates the performances under wild environmental condition evaluation sets. Table \ref{tab:table2_b} shows the performances for VoxCeleb1-based evaluation sets to investigate the adaptability of our framework.

\newpara{Environment-disentangled representation.} As observed in Table \ref{tab:table2_a}, both baseline models exhibit the lowest performances on wild environment evaluation sets, revealing vulnerabilities to the environment mismatch problem. In contrast, the models applying our framework achieve the best performances on the same evaluation sets, showing up to approximately 16\% performance improvement over the baselines. This proves that our framework's ability to extract speaker information more clearly and strengthen independence from environmental factors. The models utilising only the GRL~\cite{nam2022disentangled} also achieved performance improvements, but the more remarkable results across all evaluation sets by our framework confirm a higher capacity to exclude environmental information.

\newpara{Generalisation.} As shown in Table \ref{tab:results_main}, the models employing only the GRL~\cite{nam2022disentangled} show the highest standard deviation values across almost all evaluation sets except for one experiment with ResNet-34 on Vox1-H. Additionally, Table \ref{tab:table2_b} reveals a performance degradation on the VoxCeleb1-based evaluation sets. This highlights our claim that such direct adversarial learning-based DRL without any constraint can lead to training instability and a failure in generalisation. Conversely, our framework not only reduces the standard deviation even though employing GRL but also shows approximately 12\% improvement in the performance of baseline models, as illustrated in Table \ref{tab:table2_b}. This proves that our framework's auto-encoder effectively mitigates information loss during the DRL process, facilitating generalisation through DRL.

\section{Conclusion}

We introduce a novel adversarial learning-based DRL framework for environment robust speaker recognition. Our framework leverages an auto-encoder as a disentangler to separate a speaker representation and an environment representation from an originally entangled speaker representation of a speaker network. The auto-encoder also works to minimise unnecessary loss of vital speaker representation through reconstruction training, and consequently, the proposed framework reduces the training instability caused by adversarial learning. The proposed framework shows significant performance improvement on evaluation sets that reflect varied environments and also on standard benchmarks.

\clearpage

\ifinterspeechfinal
\section{Acknowledgements}
This work was supported by the National Research Foundation of Korea grant funded by the Korean government (Ministry of Science and ICT, RS-2023-00212845) and the ITRC (Information Technology Research Center) support program (IITP-2024-RS-2023-00259991) supervised by the IITP (Institute for Information \& Communications Technology Planning \& Evaluation).
\fi


\bibliographystyle{IEEEtran}
\bibliography{shortstrings,mybib}

\end{document}